\documentclass[12pt]{article}
\font \greekb=cmmib10 scaled \magstep1
\newcommand{\xib}{\mbox{\greekb \char 24}}
\begin{document}

\begin{center}
{\Large \bf Electromagnetic form factors and polarizations of
non-Dirac particles with rest spin 1/2} \\

\vspace{0.5 cm}

\renewcommand{\thefootnote}{*}
L.M.Slad\footnote{E-mail: slad@theory.sinp.msu.ru} \\

\vspace{0.3 cm}

{\it Skobeltsyn Institute of Nuclear Physics,
Lomonosov Moscow State University, Moscow 119991, Russian Federation}
\end{center}

\vspace{0.5 cm}

\centerline{\bf Abstract}

\begin{footnotesize}
We consider one aspect of the theoretical foundations of polarization 
experiments on elastic scattering of electrons on protons yielding form factor 
ratios incompatible with those that are extracted from nonpolarization 
experiments. We analyze the consequences of abandoning the assumption that the 
nucleon is a Dirac particle. We show that the process of elastic scattering of 
electrons on nucleons is described by the same formulas, irrespective of the 
proper Lorentz group representation associated with the nucleon as a particle 
with the rest spin 1/2.
\end{footnotesize}

\vspace{0.5 cm}

\begin{small}
\begin{center}
{\large \bf 1. Introduction}
\end{center}

A number of amazing experimental results on scattering of electrons on protons 
have recently been obtained [1]--[7]. Their totality gives evidence of an 
incompatibility between the values of the same quantity, the ratio of the 
electric form factor $G_{E}$ to the magnetic form factor $G_{M}$ of the proton,
obtained using two different methods.

One of the methods is based on extracting the form factor ratio from the 
Rosenbluth formula [8], which describes the scattering of nonpolarized 
electrons on nonpolarized protons in the laboratory reference frame,
\begin{equation}
\frac{d\sigma}{d\Omega} = \frac{\alpha^{2} E' \cos^{2}(\theta/2)}
{4 E^{3}\sin^{4}(\theta/2)} \left[
\frac{G_{E}^{2}+\tau G_{M}^{2}}{1+\tau} + 2\tau G_{M}^{2} 
\tan^{2}\frac{\theta}{2} \right] ,
\end{equation}
where $E$ and $E'$ are the respective energies of the electron in the initial 
and final states, $\theta$ is the electron scattering angle, $Q^{2}$ is the 
squared transferred momentum, $\tau = Q^{2}/4 M^{2}$, and $M$ is the proton 
mass. Recent high-precision experiments [6], [7] in the domain $Q^{2} \leq 5.5$
(GeV/c)$^{2}$ yield the $R = \mu_{p}G_{E}/G_{M}$ value (where $\mu_{p}$ is the 
magnetic moment of the proton) close to unity. This conclusion agrees with the 
outcome of Arrington's global analysis [9], to which he subjected the results 
of numerous previous experimental works.

The second method for obtaining the ratio of the electromagnetic form factors 
of the proton is based on polarization measurements. Specifically, if 
longitudinally polarized electrons are scattered on nonpolarized protons at 
rest, then the polarization of recoil protons has a transverse (in the plane of
all the particle momenta) component $P_{x}$ and the longitudinal component 
$P_{z}$, and the relation [10]
\begin{equation}
\frac{G_{E}}{G_{M}} = - \frac{P_{x}}{P_{z}} \cdot \frac{E+E'}{2M} 
\tan \frac{\theta}{2}
\end{equation}
holds, where $P_{x}$ and $P_{z}$ pertain to the rest frame of the final proton.
Using this formula, the authors of [1]--[5] concluded that there is an almost 
linear decrease in $R$ as $Q^{2}$ increases, from a value close to unity for 
small $Q^{2}$ to a value approximately equal to $0.3$ at $Q^{2} = 5.6$ 
(GeV/c)$^{2}$.

This discrepancy in the values of $R$ has caused a number of new publications 
[11]--[15] analyzing the contribution of the two-photon exchange between 
electrons and protons to the corrections to formulas (1) and (2). We note that 
radiation corrections to Rosenbluth formula (1), with the two-photon exchange
included, have been the subject of multiple investigations since times of old. 
In due time, the line was drawn here in [16], which is still used in processing
experimental data on the electron scattering on nucleons (e.g., by Arrington 
[9]). In [11]--[15], the dominant opinion is that if the structure of virtual 
hadron states in the processes involving a two-photon exchange is taken into 
account, then this reduces the discrepancy in the values of $R$ obtained from 
polarization and nonpolarization measurements several times but does not
eliminate it.

The question of the source of the discrepancies related to $R$ seems not to 
have yet obtained a clear and unambiguous theoretical resolution within the 
study of only radiation corrections. We believe it is also necessary to analyze
all aspects of the theoretical models and assumptions used in the course of 
obtaining the final result in the works on scattering of electrons on protons. 
So, conclusions regarding the polarization of recoil protons are usually 
obtained based, first, on theoretically modeling the azimuthal asymmetry [17] 
occurring because of the spin­orbit coupling in secondary scattering of protons
on a carbon target and, second, on the phenomenological 
Bargmann--Michel--Telegdi formula [18] describing the spin rotation of a 
relativistic particle in a homogeneous magnetic field. Finally, both formulas 
(1) and (2) were originally obtained under the assumption that the proton is a 
Dirac particle. This work is devoted to a thorough analysis of a number of 
consequences of dropping this assumption. We establish that formulas (1) and 
(2) are universal, i.e., they are applicable to both a Dirac and a non-Dirac 
nucleon; we thus reduce the number of theoretical aspects of polarization 
experiments on the electron scattering on protons, which, in our opinion, would
be worth comprehensively reconsidering.

\begin{center}
{\large \bf 2. Electromagnetic current and wave vectors for a non-Dirac 
particle with rest spin 1/2}
\end{center}

Assigning a nucleon the Dirac representation of the proper Lorentz group 
$L^{\uparrow}_{+}$ entirely agreed with the early views of the inner structure 
of the nucleon based on the meson model, where the physical nucleon consists of
a bare (Dirac) nucleon and a cloud of pseudoscalar pions. This assignment 
implies the most general form of the conserved nucleon electromagnetic current 
in the momentum space [19], [20]
\begin{equation}
j^{\mu}(p,p_{0}) = ie \bar{u}(p) \left[ \gamma^{\mu} F_{1}(Q^{2}) +
\frac{i\kappa}{2M} \sigma^{\mu\nu} q_{\nu} F_{2}(Q^{2}) \right] u(p_{0}),
\end{equation}
which is a polar four-vector of the orthochronous Lorentz group $L^{\uparrow}$.
Here, $\bar{u}$ and  $u$ are Dirac spinors for the nucleon, $\kappa$ is the 
anomalous magnetic moment of the nucleon expressed in nuclear magnetons, 
$q = p - p_{0}$, and $Q^{2} = -q^{2}$. The Dirac form factor $F_{1}$ and the 
Pauli form factor $F_{2}$ in (3) are related to the Sachs form
factors $G_{E}$ and $G_{M}$ in (1) and (2) as [20]
\begin{equation}
G_{E} = F_{1} - \kappa \tau F_{2}, \qquad G_{M} = F_{1} + \kappa F_{2}.
\end{equation}
The quark model of hadrons and then of quantum chromodynamics altered our 
perception of the essence of the inner structure of the nucleon. It is now 
assumed that the nucleon consists of three valence quarks and a sea of quark­
antiquark pairs and gluons. Hence, its vector-valued wave function of parton
coordinates transforms under some representation of the proper Lorentz group, a
representation belonging to the direct product of a countable set of the Dirac 
representations assigned to the quarks and a countable set of the four-vector 
representations assigned to the gluons. This product decomposes into a direct 
sum of all finite-dimensional irreducible representations of the proper Lorentz
group with half-integer spins. This conclusion concerns all nucleon states, 
both the ground state and the resonance states. Regarding the nucleon as the 
ground state, we must take into account that the nucleon has a certain spin in 
the rest frame. Therefore, each irreducible $L^{\uparrow}_{+}$-representation 
involved in describing the nucleon must contain its rest spin $1/2$. Therefore, 
in our opinion, the modern parton model of hadrons and group theory arguments
give evidence for assigning the nucleon the infinite-dimensional 
representation\renewcommand{\thefootnote}{1)}\footnote{For irreducible 
$L^{\uparrow}_{+}$-representations $\tau$, we use the notation of Gelfand and 
Yaglom [21], [22]: $\tau = (l_{0}, l_{1})$, where $2l_{0}$ is an integer and 
$l_{1}$ is an arbitrary complex number. The canonical basis of the 
representation space is related to the $SO(3)$ subgroup and is denoted by 
$\xi_{(l_{0}, l_{1})lm}$, where $l$ is the spin and $m$ is its projection on 
the third axis with $m = -l, -l+1, \ldots, l$ and  $l = |l_{0}|, |l_{0}|+1, 
\ldots$. In the general case, the last sequence is infinite. The representation
$(l_{0}, l_{1})$ is finite-dimensional, and the above sequence of spins 
terminates if $2l_{0}$ and $2l_{1}$ are integers of the same parity and 
$|l_{1}| > |l_{0}|$. The pairs $(l_{0},l_{1})$ and $(-l_{0}, -l_{1})$ describe 
the same representation. The Dirac representation is $(-1/2, 3/2) \oplus 
(1/2, 3/2)$.}
\begin{equation}
S^{3/2} = \bigoplus_{n=0}^{+\infty} \left[ \left( -\frac{1}{2}, \frac{3}{2}+n
\right) \oplus \left( \frac{1}{2}, \frac{3}{2}+n \right) \right].
\end{equation}
We already dealt with this representation in [23]--[26] in studying the theory 
of ISFIR-class fields, i.e., the fields transforming under the 
$L^{\uparrow}_{+}$-representations decomposable into an infinite direct sum of 
finite-dimensional irreducible representations. A physical implication 
established in this theory, the characteristics of its mass spectra [26], 
agrees well with the experimental picture.

We nevertheless note that all the arguments and derivations here except formula
(65) have the same force for representation (5), for any finite direct sum of 
finite-dimensional irreducible $L^{\uparrow}_{+}$-representations containing 
spin $1/2$, for any finite or infinite direct sum of infinite-dimensional 
irreducible representations of the form
\begin{equation}
S^{k_{1},{\cal K}} = \bigoplus_{n \in {\cal K}} \left[ \left( -\frac{1}{2}, 
k_{1}+n \right) \oplus \left( \frac{1}{2}, k_{1}+n \right) \right],
\end{equation}
where ${\cal K}$ is a subset of the integers and $k_{1}$ is a real number such 
that $|k_{1}| \in (0, 1/2) \cup (1/2, 1)$, and for the infinite-dimensional 
representations $S^{1/2} = (-1/2, 1/2) \oplus (1/2, 1/2)$ and 
$S^{\rm unit} = (-1/2, i\rho ) \oplus (1/2, i\rho )$, where $\rho$ is a 
positive number. If $k_{1}$ were complex in formula (6), 
$\mathop{\mathrm{Re}} k_{1} \neq 0$ and $\mathop{\mathrm{Im}} k_{1} \neq 0$, 
then a nonvanishing relativistic-invariant bilinear form and a free-field 
Lagrangian would not exist. And if we set $k_{1} = 0$ in formula (6), then the 
corresponding free-field theory would not be $CPT$ invariant [27], [28]. We 
recall that a theory of fields transforming as a finite direct sum of 
infinite-dimensional irreducible representations of the proper Lorentz group 
has several properties, listed in the introduction to [24], that make it 
unacceptable for particle physics.

If the nucleon field $\psi$ is assigned some non-Dirac representation of the 
$L^{\uparrow}_{+}$ group, then there are no reasons to restrict the Lagrangian 
to only the minimal coupling to the electromagnetic field $A_{\mu}$ and the
lowest nonminimal coupling involving the gauge-invariant antisymmetric tensor 
$F_{\mu\nu} = \partial_{\mu} A_{\nu} - \partial_{\nu} A_{\mu}$µ, as
is necessarily the case with a Dirac particle. Such a field $\psi$ in the 
general case must be assigned the electromagnetic current with a countable set 
of terms of the form
\begin{equation}
\begin{array}{l}
{\cal J}^{\mu}(p, p_{0}) = ie \left( \psi(p), \;
\left[ \Gamma^{\mu}K_{0}(Q^{2}) + \Gamma^{\mu\nu}q_{\nu} K_{1}(Q^{2}) +
\Gamma^{\mu\nu\nu_{1}}q_{\nu}q_{\nu_{1}} K_{2}(Q^{2})+ \ldots + 
\right.\right. \\
\\
\hspace{1.9cm} \left. \left. {}+\Gamma^{\mu\nu\nu_{1}\ldots \nu_{n}}q_{\nu}
q_{\nu_{1}} \ldots q_{\nu_{n}} K_{n+1}(Q^{2}) + \ldots \right] \psi(p_{0}) 
\right) , 
\end{array}  
\end{equation}
where $\Gamma^{\mu\nu}$µand µ$\Gamma^{\mu\nu\nu_{1}\ldots \nu_{n}}$, 
$n=1,2,\ldots$, are matrix tensor operators of the Lorentz group and are 
antisymmetric in $\mu$µand $\nu$. This antisymmetry ensures that the current 
${\cal J}^{\mu}$µ\ is conserved and the corresponding Lagrangian is gauge 
invariant. The needed information about the structure of any four-vector 
operator $\Gamma^{\mu}$µ\ and the details concerning the relativistic-invariant 
bilinear form of the $(\psi_{1}, \psi_{2})$ type, which we consider 
nondegenerate, can be found in [21], [22]. The description of the general 
structure of the matrix operators $\Gamma^{\mu\nu}$ and, even more so, of the 
operators $\Gamma^{\mu\nu\nu_{1}\ldots \nu_{n}}$ in (7) is still not available 
in the literature; nor is it needed for our purposes here. In what follows, we 
need the properties of tensors and tensor operators of the Lorentz group with 
respect to the spatial reflection. Unless stated otherwise, we assume that they
are everywhere the same as for the appropriate tensor product of polar 
four-vectors and polar vector operators of the orthochronous Lorentz group 
$L^{\uparrow}$.

Of all the properties of the infinite sum of operators in current (7) involved 
in describing a nonminimal electromagnetic coupling, we select only two: the 
antisymmetry of the resulting operator and its dependence on the components of 
the transferred momentum four-vector. We express this using the notation
\begin{equation}
\Lambda^{\mu\nu}(q) \equiv \Gamma^{\mu\nu} K_{1}(Q^{2}) 
+ \Gamma^{\mu\nu\nu_{1}}q_{\nu_{1}} K_{2}(Q^{2})+ \ldots 
+\Gamma^{\mu\nu\nu_{1}\ldots \nu_{n}}q_{\nu_{1}} \ldots q_{\nu_{n}} 
K_{n+1}(Q^{2})+ \ldots .
\end{equation}

In what follows, in describing the elastic scattering of electrons on nucleons,
we use the laboratory frame with the recoil nucleon momentum directed along the
third coordinate axis (the $Z$ axis) and the initial electron momentum lying in
the $XZ$ plane. Hence,
\begin{equation}
p_{0} = \{ M, 0, 0, 0 \} ,\qquad p = \{ E_{N}, 0, 0, p_{N} \} ,
\qquad q = p-p_{0} = \{q^{0}, 0, 0, q^{3} \}.
\end{equation}

In calculating the characteristics of the elastic scattering of an electron on 
a non-Dirac particle with the rest spin 1/2, we need commutation relations of 
the operators $\Lambda^{\mu\nu}(q)$ with the infinitesimal operator
$I^{12}$ of the rotation group\renewcommand{\thefootnote}{2)}\footnote{The 
information extracted from [21], [22] about Lorentz group representations and 
relativistic-invariant equations was briefly outlined in [24], but the action 
of the infinitesimal operators of the rotation group on the basis vectors 
$\xi_{\tau lm}$ was not written there, because they were not needed in that 
paper (in contrast to the present paper). We introduce the notation
$H^{3} = i I^{12}$, $H^{-} = iI^{23}+I^{31}$, and $H^{+} = iI^{23}-I^{31}$. 
Then $H^{3}\xi_{\tau lm} = m \xi_{\tau lm}$, 
$H^{-}\xi_{\tau lm} = \sqrt{(l-m+1)(l+m)} \xi_{\tau lm-1}$, and
$H^{+}\xi_{\tau lm} = \sqrt{(l-m)(l+m+1)} \xi_{\tau lm+1}$.}. In the chosen 
reference frame, they are given by
\begin{equation}
\left[ I^{12}, \Lambda^{\mu\nu}(q) \right] = -g^{1\mu}\Lambda^{2\nu}(q)+ 
g^{1\nu}\Lambda^{2\mu}(q)+g^{2\mu}\Lambda^{1\nu}(q)-
g^{2\nu}\Lambda^{1\mu}(q),
\end{equation}
where $g^{00}=-g^{11}=-g^{22}=-g^{33}=1$, $g^{\rho\sigma}=0$, and 
$\rho \neq \sigma$.

In fact, declaring that the operator $\Lambda^{\mu\nu}(q)$ is an antisymmetric 
tensor operator means that the quantity 
$(\psi_{1}, \Lambda^{\mu\nu}(q) \psi_{2})$ transforms under the group 
$L^{\uparrow}$ as an antisymmetric tensor and that the contraction 
$(\psi_{1}, \Lambda^{\mu\nu}(q) \eta_{\mu\nu} \psi_{2})$, where $\eta_{\mu\nu}$
is any antisymmetric second-rank tensor, is a relativistic-invariant 
nondegenerate bilinear form. If we assume that the field vectors $\psi_{1}$ and
$\psi_{2}$ belong to the representation space $S(g)$ of the orthochronous 
Lorentz group, then we have
\begin{equation}
[S(g)]^{-1} \Lambda^{\mu\nu}(q')\eta'_{\mu\nu} S(g)=
= \Lambda^{\mu\nu}(q)\eta_{\mu\nu}
\end{equation}
and
\begin{equation}
\begin{array}{l}
\chi'_{\mu} = {[l(g)]_{\mu}}^{\rho} \chi_{\rho}, \\
\\
\eta'_{\mu\nu} ={[T(g)]_{\mu\nu}}^{\rho\sigma} \eta_{\rho\sigma} =
\frac{1}{2} \{ {[l(g)]_{\mu}}^{\rho} {[l(g)]_{\nu}}^{\sigma} - 
{[l(g)]_{\nu}}^{\rho} {[l(g)]_{\mu}}^{\sigma} \} \eta_{\rho\sigma},
\end{array}
\end{equation}
where $\chi$ is an arbitrary polar four-vector. We consider an infinitesimal 
rotation $g_{1}$ about the third coordinate axis. Then $q' = q$, and also
\begin{equation}
\chi' = \{\chi_{0}, \;\chi _{1}-\epsilon_{12} \chi_{2},\; 
\chi_{2}+\epsilon_{12} \chi_{1}, \; \chi_{3}\} ,
\qquad S(g_{1}) = 1 + \epsilon_{12} I^{12}.
\end{equation}
Using (13), (12), and (11), we obtain commutation relations (10). Under other 
infinitesimal rotations or proper Lorentz transformations, the four-vector $q$  
changes, and formula (11) cannot lead to commutation relations of the operators
$\Lambda^{\mu\nu}(q)$ with any of the other ($\{\rho,\sigma \}\neq \{ 1, 2 \}$)
infinitesimal operators $I^{\rho\sigma}$ of the
$L^{\uparrow}_{+}$ group.

We assume that the wave function of a non-Dirac nucleon $\psi(p)$, as well as 
of a Dirac nucleon, transforming under one of the above 
$L^{\uparrow}_{+}$-representations $S_{0}$, satisfies some definite 
relativistic-invariant equation
\begin{equation}
(\Gamma^{\mu} p_{\mu} - R)\psi(p) = 0,
\end{equation}
where $R$ is a scalar matrix operator.

Because the matrix of the $\Gamma^{0}$ and $R$ operators are diagonal with 
respect to the spin $l$ and to its projection $m$ on the third axis and because
their elements are independent of $m$ [21], [22], it follows that each particle
described by some solution $\psi(p)$ of (14) in the particle rest frame can be 
assigned a definite spin and $m$-independent wave function components. In 
particular, for a spin-1/2 particle, two of its independent states in the rest 
frame can be assigned the vectors
\begin{equation}
\psi_{m}(p_{0}) = \sum_{(l_{0},l_{1}) \in S_{0}} 
u_{(l_{0},l_{1})\frac{1}{2} m}(0) \xi_{(l_{0}, l_{1})\frac{1}{2} m},  
\end{equation}
where $m=-1/2, 1/2$ and
\begin{equation}
u_{(l_{0},l_{1})\frac{1}{2} -\frac{1}{2}}(0) = 
u_{(l_{0},l_{1})\frac{1}{2} \frac{1}{2}}(0).
\end{equation}
Because (14) is invariant under spatial reflections, each particle state vector
in the particle rest frame has a certain $P$-parity, namely,
\begin{equation} 
P\psi_{m}(p_{0}) = r\psi_{m}(p_{0}), \qquad  
u_{(-\frac{1}{2},l_{1})\frac{1}{2} m}(0) = 
r u_{(\frac{1}{2},l_{1})\frac{1}{2} m}(0),
\end{equation}
where $r= +1$ or $r= -1$.

At the transition from the particle rest frame to the chosen laboratory frame 
corresponding to the element $g_{0}$ of the proper Lorentz group, the 
transformations in the space of wave vectors $\psi$ are given by the operator
$S(g_{0}) = \exp (\alpha I^{03})$, where $\tanh \alpha = v = p_{N}/E_{N}$ . 
Matrix elements of this operator are diagonal with respect to the spin 
projection on the third axis:
\begin{equation}
\exp (\alpha I^{03}) \xi_{(l_{0}, l_{1})l m} = 
\sum_{l'} A^{(l_{0},l_{1})}_{l'm,lm} (\alpha)
\xi_{(l_{0}, l_{1})l' m}.
\end{equation}
Acting with this operator on both sides of (15), we obtain the vectors
\begin{equation}
\psi_{m}(p) = S(g_{0})\psi_{m}(p_{0}) = \sum_{(l_{0},l_{1}) \in S_{0}} \sum_{l}
u_{(l_{0},l_{1})lm}(\alpha) \xi_{(l_{0}, l_{1})l m},
\end{equation}
where
\begin{equation}
u_{(l_{0},l_{1})lm}(\alpha) = A^{(l_{0},l_{1})}_{lm,\frac{1}{2} m} (\alpha)
u_{(l_{0},l_{1})\frac{1}{2} m}(0).
\end{equation}

In what follows, we use the general relations
\begin{equation}
A^{(l_{0},l_{1})}_{l -m,\frac{1}{2} -m}(\alpha) = 
A^{(-l_{0},l_{1})}_{lm,\frac{1}{2}m}(\alpha) =
\left[ A^{(l_{0},l_{1})}_{lm,\frac{1}{2}m}(-\alpha) \right]^{*}, 
\end{equation}
between matrix elements of finite transformations of the $L^{\uparrow}_{+}$ 
group. They easily follow from the result of acting by the infinitesimal 
operator $I^{03}$ on canonical basis vectors (see formulas (I.7) and 
(I.9)\renewcommand{\thefootnote}{3)}\footnote{Here and hereafter, a reference 
of the form (I.N) is to formula (N) in [24].}).

\begin{center}
{\large \bf 3. Elastic scattering cross section of a nonpolarized electron on a
nonpolarized non-Dirac particle with the rest spin 1/2}
\end{center}

If the electromagnetic current of a nucleon is given by formulas (7) and (8), 
then the elastic scattering of an electron on this nucleon with fixed values of
its spin projection on the third axis in the initial and final states 
corresponds to the matrix element
\begin{equation}
{\cal M}_{m_{2}m_{1}} = \frac{e^{2}}{q^{2}} \bar{u}_{e'}(k') \gamma_{\mu} 
u_{e}(k) \cdot \left( \psi_{m_{2}}(p), \; [\Gamma^{\mu}K_{0}(Q^{2})
+ \Lambda^{\mu\nu}(q)q_{\nu}] \psi_{m_{1}}(p_{0}) \right) ,
\end{equation}
where $k$ and $k'$ are the respective four-momenta of the incident and 
scattered electrons described by the Dirac spinors $u_{e}(k)$ and 
$\bar{u}_{e'}(k')$.

In calculating the squared modulus of the matrix element 
$\overline{{\cal M}^{2}}$ summed over the polarizations of final-state 
particles and averaged over polarizations of the initial-state particles, we 
can use well-known formulas for the electron, which is a Dirac particle. 
Specifically [29],
\begin{equation}
\frac{1}{2} \sum_{{\rm polar.}\; e,e'}[\bar{u}_{e'}(k') \gamma_{\mu} u_{e}(k)]
[\bar{u}_{e'}(k') \gamma_{\nu} u_{e}(k)]^{*} = l_{\mu\nu} =
4k_{\mu}k_{\nu} + g_{\mu\nu} + 2 (k_{\mu}q_{\nu}+k_{\nu}q_{\mu}).
\end{equation}
This equality allows eliminating the wave vectors of electrons, their Dirac 
spinors, from our calculations. But in the general case, in finding the 
quantity $\overline{{\cal M}^{2}}$, we must constantly deal with wave vectors 
of non-Dirac nucleons. To represent this quantity most simply, similar in 
structure to the form of the Dirac nucleon, we must carefully establish a 
number of general relations for elements of its electromagnetic current. We 
take up this problem now.

First, we use the relation reflecting the relativistic invariance of the 
bilinear form,
\begin{equation}
(I^{\mu\nu} \psi_{1}, \; \psi_{2}) = -(\psi_{1}, \; I^{\mu\nu} \psi_{1}),
\end{equation}
and its corollary [21], [22]
\begin{equation}
(\xi_{\tau' l' m'}, \; \xi_{\tau l m}) = \delta_{\tau'\tau^{*}} \delta_{l'l} 
\delta_{m'm} a_{\tau^{*} \tau}(l),
\end{equation}
where $\tau^{*} = (l_{0}, -l_{1}^{*}) \sim (-l_{0}, l_{1}^{*})$ if $\tau = 
(l_{0}, l_{1})$, and also relations (I.14), (I.17), (14), (15), and (19). We 
first establish the equality
\begin{equation}
(\Gamma^{\mu} \psi_{1}, \; \psi_{2}) = (\psi_{1}, \; \Gamma^{\mu} \psi_{2}),
\end{equation}
for all values of the index $\mu$, and then
\begin{equation}
\begin{array}{l}
(\psi_{m_{2}}(p), \; \Gamma^{0}\psi_{m_{1}}(p_{0})) = \delta_{m_{2}m_{1}} 
{\displaystyle \frac{1}{M}}(\psi_{m_{1}}(p), R \psi_{m_{1}}(p_{0})), \\
\\
(\psi_{m_{2}}(p), \; \Gamma^{3}\psi_{m_{1}}(p_{0})) = \delta_{m_{2}m_{1}} 
{\displaystyle \frac{q^{0}}{q^{3}M}}(\psi_{m_{1}}(p), R \psi_{m_{1}}(p_{0})).
\end{array}
\end{equation}

Second, we note that while the action of the operator $\Gamma^{0}$ on a vector 
of the canonical basis does not change the quantum number $m$ of the vector, 
the infinitesimal operators $I^{10}$ and $I^{20}$ (see relations (I.7), (I.9), 
and (I.10)) and hence the operators $\Gamma^{1}$ and $\Gamma^{2}$ found in 
accordance with (I.17) change this quantum number by $\pm 1$. Also, based on 
formula (10) and the explicit form of the operator $I^{12}$ in the canonical 
basis, it is easy to establish that the two components $\Lambda^{12}(q)$ and 
$\Lambda^{30}(q)$ of the antisymmetric tensor $\Lambda^{\mu\nu}(q)$ do not 
change the number $m$ in the vectors $\xi_{\tau l m}$, and the other components 
change it by $\pm 1$. All this, taken together with (15) and (19), where 
$m = \pm 1/2$, and with (25), yields
\begin{equation}
\begin{array}{l}
(\psi_{m_{2}}(p), U_{1}\psi_{m_{1}}(p_{0})) = \delta_{m_{2}m_{1}} 
(\psi_{m_{1}}(p),U_{1}\psi_{m_{1}}(p_{0})), \\
\\
(\psi_{m_{2}}(p), U_{2}\psi_{m_{1}}(p_{0})) = \delta_{m_{2}-m_{1}} 
(\psi_{-m_{1}}(p), U_{2} \psi_{m_{1}}(p_{0})),
\end{array}
\end{equation}
if \ $U_{1} \in \{ \Lambda^{12}(q), \Lambda^{30}(q)\}$ and $U_{2} \in 
\{\Gamma^{1},\Gamma^{2}, \Lambda^{23}(q), \Lambda^{31}(q), \Lambda^{10}(q),
\Lambda^{20}(q) \}$.

Third, we verify the intuitive expectation that two quantities
$(\psi_{m_{2}}(p), W \psi_{m_{1}}(p_{0}))$ and 
$(\psi_{-m_{2}}(p), W \psi_{-m_{1}}(p_{0}))$, where $W$ is some scalar or 
four-vector or tensor operator of the Lorentz group, have the same value, at 
least up to a phase factor. We immediately note that in a sufficiently general 
case involving elements of current (7), (8), the required proof is nontrivial. 
It is based on the fact that the parameters $\varepsilon_{\mu\nu}$ of proper 
transformations of the orthochronous Lorentz group constitute an antisymmetric 
tensor of this group. Let the field $\psi$ be a vector in the space ${\cal L}$ 
of an $L^{\uparrow}$-representation $S(g)$. The vector $\psi' = 
\exp (\varepsilon_{\mu\nu} I^{\mu\nu}/2) \psi$ belonging to the space 
${\cal L}$ can be subjected to the transformation from the group $L^{\uparrow}$
generated by $g$ in two equivalent ways, represented by the left- and right-hand
sides of the equality
\begin{equation}
S(g)\psi' = \exp\{ {[T(g)]_{\mu\nu}}^{\rho\sigma} \varepsilon_{\rho\sigma} 
I^{\mu\nu}/2\} S(g)\psi.
\end{equation}
Hence\renewcommand{\thefootnote}{4)}\footnote{Relation (30) is quite possibly 
new in group theory. A similar relation can also be given for any Lie group $G$,
taking into account that its parameters are components of vectors in the adjoint
representation space of $G$.},
\begin{equation}
S(g) \exp (\varepsilon_{\mu\nu} I^{\mu\nu}/2) =
\exp\{ {[T(g)]_{\mu\nu}}^{\rho\sigma}\varepsilon_{\rho\sigma} 
I^{\mu\nu}/2\} S(g).
\end{equation}
Let the element $g_{2}$ of the orthochronous Lorentz group correspond to 
rotation through the angle $\pi$ about the first axis (the $X$ axis) followed by
a spatial reflection. Under this transformation, which leaves the polar
four-vector $q$ unchanged, we have
\begin{equation}
\{ \eta'_{12}, \eta'_{23}, \eta'_{31}, \eta'_{01}, \eta'_{02}, \eta'_{03} \} =
\{ -\eta_{12}, \eta_{23}, -\eta_{31}, -\eta_{01}, \eta_{02}, \eta_{03} \} , 
\quad S(g_{2}) =P\exp (\pi I^{23}).
\end{equation}
In view of the relativistic invariance of the bilinear form, relations (11), 
(30), (31), and (15), the explicit form of the operator $I^{23}$ in the 
canonical basis, and relations (16) and (17), we obtain
\begin{equation}
\begin{array}{l}
( \psi_{m_{2}}(p), \; \Lambda^{\rho\sigma} (q) \eta_{\rho\sigma} 
\psi_{m_{1}}(p_{0})) = \\
\\
\hspace*{1.1cm} =( S(g_{2})\exp (\alpha I^{03}) \psi_{m_{2}}(p_{0}), \;
S(g_{2})\Lambda^{\rho\sigma} (q)\eta_{\rho\sigma} \psi_{m_{1}}(p_{0})) = \\
\\
\hspace*{1.1cm} = ( \exp (\alpha I^{03}) S(g_{2})\sum\limits_{\tau} 
u_{\tau\frac{1}{2} m_{2}}(0) \xi_{\tau\frac{1}{2} m_{2}}, \;
\Lambda^{\rho\sigma}(q) \eta'_{\rho\sigma}S(g_{2}) \sum\limits_{\tau} 
u_{\tau\frac{1}{2} m_{1}}(0) \xi_{\tau\frac{1}{2} m_{1}}) = \\
\\
\hspace*{1.1cm} =( \exp (\alpha I^{03}) P\sum\limits_{\tau} 
u_{\tau\frac{1}{2} m_{2}}(0) (-i) \xi_{\tau\frac{1}{2} -m_{2}}, \;
\Lambda^{\rho\sigma}(q) \eta'_{\rho\sigma}P \sum\limits_{\tau} 
u_{\tau\frac{1}{2} m_{1}}(0) (-i) \xi_{\tau\frac{1}{2} -m_{1}}) = \\
\\
\hspace*{1.1cm} =( \exp (\alpha I^{03}) P\sum\limits_{\tau} 
u_{\tau\frac{1}{2} -m_{2}}(0) \xi_{\tau\frac{1}{2} -m_{2}}, \;
\Lambda^{\rho\sigma}(q) \eta'_{\rho\sigma}P \sum\limits_{\tau} 
u_{\tau\frac{1}{2} -m_{1}}(0) \xi_{\tau\frac{1}{2} -m_{1}}) = \\
\\
\hspace*{1.1cm} =( \exp (\alpha I^{03})r \psi_{-m_{2}}(p_{0}), \;
\Lambda^{\rho\sigma} (q) \eta'_{\rho\sigma} r \psi_{-m_{1}}(p_{0})) = \\
\\
\hspace*{1.1cm} =( \psi_{-m_{2}}(p), \; \Lambda^{\rho\sigma} (q) 
\eta'_{\rho\sigma} \psi_{-m_{1}}(p_{0})) .
\end{array}
\end{equation}
Hence, the relation
\begin{equation}
(\psi_{m_{2}}(p), \; \Lambda^{\mu\nu} (q) \psi_{m_{1}}(p_{0})) =
\kappa_{1} (\psi_{-m_{2}}(p), \; \Lambda^{\mu\nu} (q) \psi_{-m_{1}}(p_{0}))
\end{equation}
holds, where $\kappa_{1} = 1$ if neither of the two indices $\mu$ and $\nu$ is 
equal to unity and $\kappa_{1} = -1$ otherwise. Similarly, we find that
\begin{equation}
(\psi_{m_{2}}(p), \; \Gamma^{\mu} \psi_{m_{1}}(p_{0})) =
\kappa_{0} (\psi_{-m_{2}}(p), \; \Gamma^{\mu} \psi_{-m_{1}}(p_{0})),
\end{equation}
where $\kappa_{0} = 1$ if $\mu = 0$, $2$ or $3$ and $\kappa_{0} = -1$ if 
$\mu = 1$.

Fourth, using relations (28), (I.15), (10), (24), (15), and (19) and the 
explicit form of the infinitesimal operator $I^{12}$, we obtain
\begin{equation}
\begin{array}{l}
(\psi_{-m_{1}}(p), \; \Gamma^{2} \psi_{m_{1}}(p_{0})) =
2im_{1} (\psi_{-m_{1}}(p), \; \Gamma^{1} \psi_{m_{1}}(p_{0})), \\
\\
(\psi_{-m_{1}}(p), \; \Lambda^{2\sigma} \psi_{m_{1}}(p_{0})) =
2im_{1} (\psi_{-m_{1}}(p),\; \Lambda^{1\sigma} \psi_{m_{1}}(p_{0})),
\end{array}
\end{equation}
where $\sigma = 0$ or $\sigma = 3$.

Only now, based on relations (22), (23), (27), (28), and (33)--(35), do we find 
the simplest expression for the squared modulus of the matrix element 
$\overline{{\cal M}^{2}}$ describing the process of elastic scattering of 
nonpolarized electrons on nonpolarized nucleons. Next, using the standard 
formula (see, e.g., [30]) expressing the differential cross section of the 
process in terms of $\overline{{\cal M}^{2}}$, we obtain a distribution over the
solid angle of scattered electrons. We verify that this distribution is 
described by exactly Rosenbluth formula (1). The role of the electromagnetic 
form factors $G_{E}$ and $G_{M}$ is then played by the quantities
\begin{equation}
\hspace*{-1.1cm} G_{E} = \frac{C}{\sqrt{\tau +1}} 
(\psi_{+1/2}(p), \; [K_{0}(Q^{2})R -Mq^{3}\Lambda^{03}(q)] \psi_{+1/2}(p_{0})),
\end{equation}
\begin{equation}
G_{M} = \frac{M C}{\sqrt{\tau}} 
(\psi_{+1/2}(p), \; [K_{0}(Q^{2})\Gamma^{1}+q^{0}\Lambda^{10}(q)-
q^{3}\Lambda^{13}(q)] \psi_{-1/2}(p_{0})) ,
\end{equation}
where
\begin{equation}
C = (\psi_{+1/2}(p_{0}), R \psi_{+1/2}(p_{0}))^{-1}.
\end{equation}

\begin{center}
{\large \bf 4.Polarization of recoil nucleons in elastic scattering of polarized
electrons on nonpolarized non-Dirac nucleons}
\end{center}

The calculations taking the polarization of an electron with the mass $m_{e}$ 
into account are based on the relation [29]
\begin{equation}
u_{e}(k) \bar{u}_{e}(k) = \frac{1}{2} (\hat{k}+m_{e})(1-\gamma^{5}\hat{a}),
\end{equation}
where the components of the axial four-vector $a$ are
\begin{equation}
a^{0} = \frac{{\bf k}\xib}{m_{e}}, \qquad {\bf a} = \xib +
\frac{({\bf k}\xib){\bf k}}{m_{e}(m_{e}+E)}
\end{equation}
with $\xib$ being the unit vector of the electron polarization in the rest frame
of the electron. We recall that the polarization of a spinor particle in some 
state is twice the average value of spin in this state.

Summing over the polarizations of the scattered relativistic electron, using 
formulas (39) and (40), and neglecting terms of the order of $m_{e}/E$, one 
obtains [10]
\begin{equation}
\sum_{{\rm polar.}\; e'}[\bar{u}_{e'}(k') \gamma_{\mu} u_{e}(k)]
[\bar{u}_{e'}(k') \gamma_{\nu} u_{e}(k)]^{*} = l_{\mu\nu} +
2i\xi_{\|} \varepsilon_{\mu\nu\rho\sigma} q^{\rho} k^{\sigma},
\end{equation}
where $l_{\mu\nu}$ is given by relation (23) and $\xi_{\|}$ denotes the 
longitudinal polarization of the initial-state electron,
$-1 \leq \xi_{\|} \leq +1$. We note that the transverse polarization of the 
initial electron, being nonzero as long as $|\xi_{\|}| \neq 1$, makes a 
negligibly small contribution ${\cal O}(m_{e}/E)$ to the right-hand side of 
(41).

The analogue of (39) and (40) is an essential element in the standard procedure 
for calculating the polarization of Dirac particles produced in some process 
[29]. It was used in [10] for recoil nucleons in the scattering of polarized 
electrons on a nonpolarized target. Considering non-Dirac particles, we lose 
this procedure together with formulas (39) and (40).

Our method for finding the polarization of recoil nucleons is based on the 
explicit form of their state vectors $\Psi_{m}^{\lambda'}$, each of which is a 
superposition of two state vectors (19) with definite spin projections on the
third axis, and the coefficients in this superposition are given by the 
appropriate matrix elements of the electron scattering on nucleons, namely,
\begin{equation}
\Psi_{m_{1}}^{\lambda'} = N_{m_{1}}^{\lambda'} \left[ 
{\cal M}_{-\frac{1}{2} m_{1}}^{\lambda'} 
\psi_{-\frac{1}{2}}(p)+ {\cal M}_{\frac{1}{2} m_{1}}^{\lambda'}
\psi_{\frac{1}{2}}(p)) \right] ,
\end{equation}
where
\begin{equation} 
N_{m_{1}}^{\lambda'} =(|{\cal M}_{-\frac{1}{2} m_{1}}^{\lambda'}|^{2}+
|{\cal M}_{\frac{1}{2} m_{1}}^{\lambda'}|^{2})^{-1/2}
\end{equation} 
and ${\lambda'}$ is the polarization (helicity) of the scattered electron (it is
not written explicitly in formula (22)).

The average value $b$ of a quantity characterizing recoil nucleons and described
by an operator $B$ is found by the formula
\begin{equation}
b = D_{0}^{-1} \sum_{\lambda', m_{1}}
(\Psi_{m_{1}}^{\lambda'}, \; B \Psi_{m_{1}}^{\lambda'}) w_{m_{1}}^{\lambda'},
\end{equation}
where
\begin{equation}
D_{0} = (\psi_{m}(p_{0}), \; \psi_{m}(p_{0}))
\end{equation}
and $w_{m_{1}}^{\lambda'}$ is the probability that the created nucleon and the 
electron have one of the two possible values of the spin projection $m_{1}$ on 
the third axis and one of the two possible values of polarization (helicity) 
$\lambda'$. Obviously,
\begin{equation}
w_{m_{1}}^{\lambda'} = N^{2} (N_{m_{1}}^{\lambda'})^{-2},  
\end{equation}
where
\begin{equation}
N = ( \sum_{\lambda', m_{1}, m_{2}}  
|{\cal M}_{m_{1} m_{2}}^{\lambda'}|^{2} )^{-1/2}.
\end{equation}

In the relativistic quantum theory, the spin operator is identified with the 
antisymmetric tensor operator $i I^{\mu\nu}$, whose components are generators 
of the proper Lorentz group. We are primarily interested in the average values 
of purely spatial components of spin, constituting a vector ${\bf s}$. It is 
essential that in the rest frame of a Dirac or non-Dirac spinor particle, for 
any state with a definite $P$-parity, the average values of the other 
components of the spin tensor are equal to zero:
\begin{equation}
(\psi(p_{0}), \; I^{0i} \psi(p_{0})) = 0,
\end{equation}
where
\begin{equation}
\psi(p_{0}) = c_{-1/2}\psi_{-1/2}(p_{0}) + c_{1/2}\psi_{1/2}(p_{0}),
\end{equation}
$i=1,2,3$, and $c_{m}$ are arbitrary superposition coefficients. Indeed, using 
relations (49) and (17), the invariance of the bilinear form under spatial 
relations, and relation (11) with the replacement of $\Lambda^{\mu\nu}(q)$ with 
$I^{\mu\nu}$ and $S(g)$ with $P^{-1}$, we obtain
\begin{equation}
\begin{array}{l}
(\psi(p_{0}), \; [I^{0i}\eta_{0i}+I^{jk}\eta_{jk}] \psi(p_{0})) = 
(P \psi(p_{0}), \; [I^{0i}\eta_{0i}+I^{jk}\eta_{jk}] P \psi(p_{0}))= \\
\\
\hspace{5.2cm} = (\psi(p_{0}), \; P^{-1} [I^{0i}\eta_{0i}+I^{jk}\eta_{jk}] 
P \psi(p_{0}))= \\
\\
\hspace{5.2cm} =(\psi(p_{0}), \; [-I^{0i}\eta_{0i}+I^{jk}\eta_{jk}] 
\psi(p_{0})).
\end{array}
\end{equation}

Interestingly, Frenkel, who suggested a relativistic description of the 
classical spin as an antisymmetric tensor prior to the appearance of the Dirac 
equation, formulated the rule that in the rest frame of a particle, only purely 
spatial components of its spin tensor can be nonzero [31] (a review of 
Frenkel's works on the classical theory of spin can be found in [32]).

Using relation (48) and the known formulas for a Lorentz transformation of the 
components of an antisymmetric tensor in passing from one inertial reference 
frame to another, we obtain
\begin{equation}
s_{x} = \cosh \alpha \cdot s_{0x}, \qquad s_{y} = \cosh \alpha \cdot s_{0y},
\qquad s_{z} = s_{0z},
\end{equation}
where the vector ${\bf s}_{0}$ refers to the rest frame of the final nucleon 
and the vector ${\bf s}$ refers to our chosen laboratory frame.

Consecutively replacing the operator $B$ in formula (44) with the operators 
$i I^{23}$, $i I^{31}$, and $i I^{12}$, recalling their action on the canonical 
basis vectors, and taking (42), (45), (46), (19)--(21), and (17) into account, 
we obtain
\begin{equation}
\begin{array}{l}
s_{x} = N^{2} \sum\limits_{\lambda', m_{1}} {\rm Re} 
\left[ ({\cal M}_{\frac{1}{2} m_{1}}^{\lambda'})^{*}
{\cal M}_{-\frac{1}{2} m_{1}}^{\lambda'} \right] D(\alpha ), \\
\\
s_{y} = N^{2} \sum\limits_{\lambda', m_{1}} {\rm Im} 
\left[ ({\cal M}_{\frac{1}{2} m_{1}}^{\lambda'})^{*}
{\cal M}_{-\frac{1}{2} m_{1}}^{\lambda'} \right] D(\alpha ), \\
\\
s_{z} =  \frac{N^{2}}{2} \sum\limits_{\lambda', m_{1}} 
\left[ |{\cal M}_{\frac{1}{2} m_{1}}^{\lambda'})|^{2}-
|{\cal M}_{-\frac{1}{2} m_{1}}^{\lambda'}|^{2} \right],
\end{array}
\end{equation}
where
\begin{equation}
D(\alpha ) = D_{0}^{-1} \sum_{\tau',\tau \in S_{0}} \sum_{l}
\left( u_{\tau'l\frac{1}{2}}(\alpha) 
\xi_{\tau'l\frac{1}{2}}, \; (l+1/2)
u_{\tau l-\frac{1}{2}}(\alpha) \xi_{\tau l\frac{1}{2}}\right). 
\end{equation}

Calculations based on (52), (22), (41), (23), (27), (28), and (33)--(37) give 
the results
\begin{equation}
\begin{array}{l}
s_{x} = 8 N^{2} \xi_{\|} \left( {\displaystyle \frac{e^{2}}{q^{2}}}\right)^{2}
{\displaystyle \frac{\tau}{C^{2}}} \sqrt{{\displaystyle \frac{\tau}{1+\tau}}} 
\cot{\displaystyle \frac{\theta}{2}} 
G_{E} G_{M} D(\alpha), \\
\\
s_{y} = 0, \\
\\
s_{z} = - 4 N^{2} \xi_{\|} \left( {\displaystyle\frac{e^{2}}{q^{2}}}\right)^{2}
{\displaystyle \frac{(E+E')\tau}{MC^{2}}} 
\sqrt{{\displaystyle \frac{\tau}{1+\tau}}} G_{M}^{2}.
\end{array}
\end{equation}
Using formulas (51) and expressing the polarization vector in terms of the rest
spin of a fermion (${\bf P} = 2{\bf s}_{0}$), we hence obtain
\begin{equation}
\frac{G_{E}}{G_{M}} = - \frac{P_{x}}{P_{z}} \cdot \frac{E+E'}{2M} 
\cdot \frac{\cosh \alpha}{D(\alpha)} \tan\frac{\theta}{2}.
\end{equation}

We now prove that the equality
\begin{equation}
D(\alpha) =\cosh \alpha 
\end{equation}
holds for any of the $L^{\uparrow}_{+}$-representations $S_{0}$ that we 
consider.

We introduce a four-vector operator $L^{\mu}$ such that it has a single 
coupling to each irreducible representation $(\pm 1/2, l_{1})$ belonging to 
$S_{0}$, namely, to the representation $(\mp 1/2, l_{1})$. We set all the 
arbitrary constants assigned to the operator $L^{0}$ in the general case equal 
to unity. Then (see formulas (I.20) and (I.21))
\begin{equation}
L^{0} \xi_{(\pm \frac{1}{2}, l_{1})lm} = (l+ 1/2) 
\xi_{(\mp \frac{1}{2}, l_{1})lm}.
\end{equation}
We also introduce a polar four-vector $\zeta$ such that only its time component
is nonzero in the chosen laboratory frame: $\zeta_{0} = 1$, $\zeta_{i} = 0$, 
$i=1,2,3$. In the rest frame of the final-state nucleon, passing to which 
corresponds to the element $g_{0}^{-1}$ of the $L^{\uparrow}_{+}$ group, we 
have
\begin{equation}
\zeta' = \{ \zeta'_{0},\; \zeta'_{1},\; \zeta'_{2},\; \zeta'_{3} \}  = 
\{ \cosh \alpha, \; 0, \; 0, \;  \sinh \alpha \}.
\end{equation}

Turning to relations (53), (57), (25), (16), (20), and (21) with the 
relativistic invariance of the bilinear form and relations (I.13) and (58) 
taken into account, we obtain the chain of equalities
\begin{equation}
\begin{array}{l}
D(\alpha) = D_{0}^{-1} \sum\limits_{l'_{0},l'_{1}} \sum\limits_{l_{1}} 
\sum\limits_{l_{0}= -1/2}^{1/2} \sum\limits_{l',l} 
\left( u_{(l'_{0},l'_{1})l\frac{1}{2}}(\alpha) 
\xi_{(l'_{0},l'_{1})l'\frac{1}{2}}, 
\; (l+1/2)r u_{(-l_{0},l_{1})l\frac{1}{2}}(\alpha)
\xi_{(l_{0},l_{1})l\frac{1}{2}} \right) = \\
\hspace{1.0cm} = D_{0}^{-1} r\sum\limits_{l'_{0},l'_{1}} \sum\limits_{l_{1}} 
\sum\limits_{l_{0}= -1/2}^{1/2} \sum\limits_{l',l} 
\left( u_{(l'_{0},l'_{1})l\frac{1}{2}}(\alpha) 
\xi_{(l'_{0},l'_{1})l'\frac{1}{2}}, \; L^{\mu}\zeta_{\mu}
u_{(-l_{0},l_{1})l\frac{1}{2}}(\alpha)
\xi_{(-l_{0},l_{1})l\frac{1}{2}}\right) = \\
\\
\hspace{1.0cm} = D_{0}^{-1} r(\psi_{m}(p), \; L^{\mu}\zeta_{\mu} 
\psi_{m}(p)) = \\
\\
\hspace{1.0cm} = D_{0}^{-1} r(\psi_{m}(p_{0}), \; S(g_{0}^{-1}) 
L^{\mu}\zeta_{\mu} S^{-1}(g_{0}^{-1}) \psi_{m}(p_{0})) = \\
\\
\hspace{1.0cm} = D_{0}^{-1} r(\psi_{m}(p_{0}), \; L^{\mu} \zeta'_{\mu} 
\psi_{m}(p_{0})) = \\
\\
\hspace{1.0cm} = D_{0}^{-1} r(\psi_{m}(p_{0}), \; [\cosh \alpha \cdot L^{0} + 
\sinh \alpha \cdot L^{3}] \psi_{m}(p_{0})).
\end{array}
\end{equation}
In addition, the next-to-last relation in (59) and formula (17), taken together
with the invariance of the bilinear form under spatial reflections, relation 
(I.13) with $S(g) = P^{-1}$, and Eq. (58), imply that
\begin{equation}
\begin{array}{l}
D(\alpha) = D_{0}^{-1} r(P\psi_{m}(p_{0}), \; L^{\mu} \zeta'_{\mu} 
P\psi_{m}(p_{0})) = \\
\\ 
\hspace{1.0cm} = D_{0}^{-1} r(\psi_{m}(p_{0}), \; 
P^{-1}L^{\mu} \zeta'_{\mu} P\psi_{m}(p_{0})) = \\
\\
\hspace{1.0cm} = D_{0}^{-1} r(\psi_{m}(p_{0}), \; [L^{0} \zeta'_{0}- L^{i} 
\zeta'_{i}] \psi_{m}(p_{0})) = \\
\\
\hspace{1.0cm} = D_{0}^{-1} r(\psi_{m}(p_{0}), \; 
[\cosh \alpha \cdot L^{0} - \sinh \alpha \cdot L^{3}] \psi_{m}(p_{0})).
\end{array}
\end{equation}
Comparing the final results in chains (59) and (60), we have
\begin{equation}
(\psi_{m}(p_{0}), \; L^{3} \psi_{m}(p_{0})) = 0.
\end{equation}
Therefore, recalling (45), we obtain
\begin{equation}
\begin{array}{l}
D(\alpha) = \cosh \alpha \cdot D_{0}^{-1} r(\psi_{m}(p_{0}), \; 
L^{0} \psi_{m}(p_{0})) = \\
\\
\hspace{1.0cm} = \cosh \alpha \cdot D_{0}^{-1} \sum\limits_{l'_{0},l'_{1}} 
\sum\limits_{l_{1}} \sum\limits_{l_{0}= -1/2}^{1/2} 
\left( u_{(l'_{0},l'_{1})\frac{1}{2}\frac{1}{2}}(0) 
\xi_{(l'_{0},l'_{1})\frac{1}{2}\frac{1}{2}}, \; 
ru_{(l_{0},l_{1})\frac{1}{2}\frac{1}{2}}(0)
\xi_{(-l_{0},l_{1})\frac{1}{2}\frac{1}{2}}\right) = \\
\hspace{1.0cm} = \cosh \alpha \cdot D_{0}^{-1} \sum\limits_{l'_{0},l'_{1}} 
\sum\limits_{l_{1}} \sum\limits_{l_{0}= -1/2}^{1/2} 
\left( u_{(l'_{0},l'_{1})\frac{1}{2}\frac{1}{2}}(0) 
\xi_{(l'_{0},l'_{1})\frac{1}{2}\frac{1}{2}}, \; 
u_{(-l_{0},l_{1})\frac{1}{2}\frac{1}{2}}(0)
\xi_{(-l_{0},l_{1})\frac{1}{2}\frac{1}{2}}\right) = \\
\\
\hspace{1.0cm} = \cosh \alpha \cdot D_{0}^{-1} (\psi_{m}(p_{0}), \; 
\psi_{m}(p_{0})) = \cosh \alpha,
\end{array}
\end{equation}
which was to be proved.

Therefore, the same formula (2) (obtained previously for Dirac nucleons) holds 
regardless of which non-Dirac representation $S_{0}$ describes the nucleon.

We consider the particular case of equality (56) where the representation 
$S_{0}$ is the direct sum $\tau^{*} \oplus \tau = (-1/2, 1/2+n) \oplus 
(1/2, 1/2+n)$ of two irreducible finite-dimensional representations of the 
proper Lorentz group, where $n$ is any positive integer. We can set [24]
\begin{equation}
a_{\tau^{*}\tau}(l) = (-1)^{l-1/2}.
\end{equation}
In accordance with (16) and (17), we fix the values of the components of the 
state vectors $\psi_{m}(p_{0})$ as
\begin{equation}
u_{(\frac{1}{2},\frac{1}{2}+n)\frac{1}{2}\frac{1}{2}}(0) =
u_{(\frac{1}{2},\frac{1}{2}+n)\frac{1}{2}-\frac{1}{2}}(0) =
r u_{(-\frac{1}{2},\frac{1}{2}+n)\frac{1}{2}\frac{1}{2}}(0) =
r u_{(-\frac{1}{2},\frac{1}{2}+n)\frac{1}{2}-\frac{1}{2}}(0) = 1/\sqrt{2}.
\end{equation}
Then $D_{0} = r$, and relation (56) with (53), (45), (25), (63), (20), and (21)
taken into account becomes
\begin{equation}
\sum_{l=1/2}^{n-1/2} (-1)^{l-1/2} (l+1/2)
\left[ \left| 
A_{l\frac{1}{2},\frac{1}{2}\frac{1}{2}}^{(\frac{1}{2},\frac{1}{2}+n)}
(\alpha)\right|^{2} + \left|
A_{l\frac{1}{2},\frac{1}{2}\frac{1}{2}}^{(\frac{1}{2},\frac{1}{2}+n)}
(-\alpha)\right|^{2} \right] = 2 \cosh \alpha.
\end{equation}
We first discovered and verified this identity with $n = 1, 2, \ldots, 5$ for 
finite transformations of the proper Lorentz group directly, knowing the 
explicit form of the appropriate matrix elements.

\begin{center}
{\large \bf 5. Concluding remarks}
\end{center}

The obtained results concerning the polarizations of a non-Dirac particle with 
the rest spin 1/2 are rather counterintuitive. Our original expectations were 
based on the fact that the state vector of such a particle, referred to the 
$L^{\uparrow}_{+}$-representation $S^{3/2}$ given by Eq. (5), for example, 
should have all half-integer spin values in the laboratory frame. It would seem
that the average value of the spin projection onto any direction noncoincident 
with the particle momentum direction, also being the same in both the rest 
frame and the laboratory frame, should depend on both the particle velocity and
the specific distribution of the components of its state vector (15) over the 
irreducible representations involved in $S^{3/2}$. These a priori statements 
are nullified by the remarkable relation (56), whose existence could not have 
been foreseen without a detailed analysis.

With this paper, we have fully clarified one aspect of the theoretical 
foundations of experiments on elastic scattering of electrons on protons by 
establishing that the freedom inherent in assigning one proper Lorentz group 
representation or another to the nucleon has no relation whatsoever to the 
contradiction encountered in the results of polarization and nonpolarization 
experiments, the results gleaned using the respective formulas (2) and (1).

Elsewhere, we may perhaps sketch a number of problems relating to another 
aspect of the theoretical foundations of polarization experiments [1]--[5]: the
validity of the Bargmann--Michel--Telegdi equation for the spin rotation of a 
relativistic nucleon in its motion in a homogeneous magnetic field. The 
magnetic field between two targets is a necessary ingredient for measuring the 
longitudinal component of the recoil proton polarization because the secondary 
proton scattering itself allows extracting only the transverse polarization
components. But because of the spin rotation, the longitudinal polarization 
component is manifested as a certain change in the transverse component $P_{x}$
(in the plane of the momenta of the recoil proton and electrons) and as the 
appearance of the $P_{y}$ component (in the direction perpendicular to the 
above plane), which would have been zero after the scattering of an electron 
(see the second equality in (54)).

{\bf Acknowledgments}. The author is deeply grateful to S.P. Baranov and 
V.E. Troitsky for the numerous stimulating discussions of the problems 
considered in this paper.

\end{small}
\end{document}